\documentclass[conference,10pt]{IEEEtran}
\IEEEoverridecommandlockouts
\usepackage{cite}
\usepackage{amsmath,amssymb,amsfonts,mathrsfs}
\usepackage{algorithmic}
\usepackage{graphicx}
\usepackage{textcomp}
\usepackage{braket}
\def\BibTeX{{\rm B\kern-.05em{\sc i\kern-.025em b}\kern-.08em
    T\kern-.1667em\lower.7ex\hbox{E}\kern-.125emX}}

\UseRawInputEncoding

\begin{document}

\title{Canonical Thermodynamics
\\
}

\author{Arnaldo Spalvieri \\
Dipartimento di Elettronica, Informazione e Bioingegneria \\
Politecnico di Milano\\
arnaldo.spalvieri@polimi.it \\
ORCID 0000-0002-8336-7996
}

\maketitle

\begin{abstract}

The paper demonstrates that the canonical probability distribution
of the occupancy numbers of a bosonic system is multinomial, and
shows how the thermodynamics of the canonical system descends from
this distribution. The categorical distribution (i.e. the
one-particle probability distribution of occupancy of the quantum
eigenstates allowed to a particle of the system) of the
multinomial distribution should be derived from constrained
maximization of the Shannon entropy of the multinomial
distribution. However, since the multinomial distribution
intractable, one must renounce to a closed-form solution to the
constrained maximization problem. The analysis is then focused on
the thermal state, that is characterized by the constraint on
system's expected energy. In this case, the paper proposes to
consider a suboptimal tractable categorical distribution, which is
likely to be close to the actual categorical maximizer, and shows
that the one-particle Boltzmann distribution is a good
approximation to the actual categorical maximizer only in certain
cases, including the classical regime. The unexpected result is
that, in the general case, the approximation that we find to the
categorical maximizer is not of exponential type, or, in other
words, is not the one-particle Boltzmann distribution. As a
consequence, the probability distribution of microstates is not of
exponential type. As in the standard analysis, it is always equal
to the product of factors but, in the general case, these factors
are not the Boltzmann factors, therefore the probability of a
microstate can be different from the probability of another
microstate even when the two have the same energy.

\vspace{0.5cm}

{\em Keywords:}\hspace{0.2cm}{\bf Thermodynamics, Canonical
Systems, Occupancy Numbers, Multinomial Distribution, Canonical
Typicality, Maximum Entropy Principle, Gibbs Correction Factor,
Sackur-Tetrode Entropy Formula.}

\end{abstract}

\section{Introduction}

From the times of Boltzmann and Gibbs there is a long debate,
still today ongoing, about the identification of the microscopic
entropy of a closed system, e.g. Boltzmann's entropy or the
Shannon-Gibbs entropy with or without the Gibbsian correction term
$-\log(N!)$, where $N$ is the number of particles of the system,
with the macroscopic entropy of the system as it is defined in
thermodynamics (Clausius' entropy).

The microcanonical entropy, or Boltzmann's entropy, applies  only
to systems that can occupy a set of equally probable quantum
states, which is a too strong limitation. The famous Gibbs paradox
shows that also the Shannon-Gibbs entropy of the Boltzmann
distribution of microstates of canonical systems cannot be
identified with the thermodynamic entropy. In fact, Gibbs
introduced the correction term $\log(N!)$ to the entropy of
microstates, to make it consistent, at least in the classical
case, with the thermodynamic entropy. When replaced by the first
two terms of Stirling's asymptotic expansion
\begin{equation}\log(N!)= N\log(N)-N+ 0.5 \log(2 \pi N) + {\cal O}(N^{-1}), \label{stirling}
\end{equation} the Gibbsian correction term becomes the fundamental ingredient of the
Sackur-Tetrode formula for the thermodynamic entropy. However, the
Sackur-Tetrode entropy formula holds only in the classical regime,
as it is apparent from the fact that it can become negative at low
temperature. Besides this limitation, it should also be observed
that the term $-\log(N!)$ is a sort of {\em deus ex machina} whose
introduction has been justified in the past by the difference
between distinguishable particles and indistinguishable particles,
but it is questionable that this difference should actually have
any impact on the physical properties of the system, see e.g.
\cite{swe2002}. In the end, the term $-\log(N!)$ is still today
object of debate and discussion, see for instance
\cite{sasa,tasaki,murashita}. More generally, still today the
entropy of canonical systems is object of ongoing research, see
for instance the recent paper \cite{psapprox}.

The picture changes radically if, in place of canonical systems,
we consider a grand-canonical system, i.e. an open system  that
randomly exchange particles along time with the surrounding
environment at the equilibrium with it. More than one century ago,
Bose and Einstein derived from the principle of constrained
entropy maximization the probability distribution of the occupancy
numbers, and, from it, the thermodynamic entropy of the photon gas
in \cite{bose} and of the gas of massive particles in
\cite{einstein24}. If we regard the occupancy numbers as a
macroscopic manifestation of the system, as it actually is being
they a sum over all the particles of the system, we realize that
the Bose-Einstein approach resolves the conflict between
microscopic and macroscopic entropy in a really brilliant way: it
knocks out of physics the microscopic entropy. However, the
particles of a canonical system do not join and leave it,
therefore, although the Bose-Einstein approach can lead to very
good approximations also in the case of closed systems, strictly
speaking it does not apply, for instance, to a gas of massive
particles confined inside a container.



To sum up, while it is universally recognized that the entropy of
a grand canonical system is the Shannon entropy of his occupancy
numbers multiplied by Boltzmann's constant, still today scientists
have not reached wide agreement about the entropy of canonical
systems. Basically, this difficulty arises from the difficulty of
finding the probability distribution of the occupancy numbers of
canonical systems. Today, the most recognized approach is that of
chapter 9 of \cite{reif}, where it is postulated that the sought
distribution is proportional to the product of the Boltzmann
factors. The distribution proposed in \cite{reif} generated a vast
body of literature, see for instance
\cite{borrmann,borrmann2,gottlieb,giraud,barghathi}. The extensive
bibliography of \cite{barghathi} bears witness to the effort made
by the scientific community to advance the understanding of the
properties of the distribution proposed in \cite{reif}. However,
we will show in this paper that this distribution does not
represent a canonical system. This observation, together with the
still unsolved puzzle of the Gibbsian correction term, motivates
us to find the probability distribution of the occupancy numbers
of the canonical system and to show that, when the system is at
the thermal state, its Shannon entropy can be identified with
system's thermodynamic entropy.

%

\section{Occupancy numbers of a system with fixed and
known number of bosons}

Consider a system of $N$ particles of the same species, let ${\cal
C}=\{1,2, \cdots,|{\cal C}|\}$, be the set of quantum eigenstates
allowed to a particle, let $c_i \in {\cal C}, \ i=1,2, \cdots, N,$
represent the generic eigenstate of the $i$-th particle, and let
${s}=(c_1,c_2,\cdots,c_N)$, ${s} \in {\cal S}={\cal C}^N$,
represent the generic eigenstate of the system (the so-called {\em
microstate} in many textbooks, e.g. \cite{pathria}). The number of
particles that occupy eigenstate $c$ is
\begin{align} n_c({s})=\sum_{i=1}^N \delta_{c,c_i},  \nonumber \end{align} where
\begin{align} \delta_{x,y}
=\left\{
\begin{array}{cc}  1, &
 \mbox{if} \ x=y,\\
0,  &  \mbox{elsewhere}.
\end{array} \right.
\nonumber\end{align} In the following, when unnecessary we will
omit the dependance of the occupancy number on ${s}$ and we will
denote $n$,
\[{n}=(n_1,n_2, \cdots, n_{|{\cal C}|}),\]
the vector of the occupancy numbers, or, in short, the occupancy
vector.  The size $|\cal{N}|$ of the set $\cal{N}$ of the
occupancy vectors allowed to a system with $N$ particles is
\[|{\cal N}|=\frac{(N+|{\cal C}|-1)!}{N!(|{\cal C}|-1)!}.\]
Let the subset ${\cal S}_{{n}}$ of ${\cal S}$ be the set of all
the vectors ${{s}}$ that have the same vector ${n}({{s}})$ of
occupancy numbers. ${\cal S}_{{n}}$ contains only and all the
$W_{{n}}$ permutations of anyone of its vectors, where
 \begin{align} W_{{n}} =\left\{
\begin{array}{cc}  \frac{N!}{\prod_c
n_c!}, &
 \mbox{if} \ \sum_{c}n_c=N,\\
0,  &  \mbox{elsewhere},
\end{array} \right.
\nonumber \end{align} is the multinomial coefficient. When ${n}$
spans the entire set ${\cal N}$, the set $\{{\cal S}_{{n}}\}$ form
a disjoint partition of ${\cal S}$,
\[\bigcup_{{n}\in {\cal N}}{\cal
S}_{{n}}={\cal S}, \ \ \bigcap_{{n}\in {\cal N}}{\cal
S}_{{n}}=\emptyset,\] hence each $s \in {\cal S}$ belongs to one
and only one of the elements of $\{{\cal S}_n\}$.

\section{Canonical density operator in the symmetric Hilbert subspace}

Let the universe consist of $U$ of bosons, denote its Hilbert
space ${\cal H}_{\cal U}={\cal H}_{\cal C}^{\otimes U}$, where
$\otimes$ denotes the Kronecker product and ${\cal H}_{\cal C}$ is
the Hilbert space of the quantum states allowed to one boson, and
consider the restriction of the Hilbert space ${\cal H}_{{\cal
U},\epsilon}\subseteq{\cal H}_{\cal U}$ spanned by the eigenstates
belonging to the strongly typical set ${\cal T}_{{\cal
U},\epsilon}$,
\begin{align} {\cal T}_{{\cal U},\epsilon}=  \left\{\tau:
\left|\frac{n_c(\tau)}{U}-P_c\right|< \epsilon, \ \forall c \in
{\cal C} \right\},\label{typset}
\end{align}
where $\epsilon>0$ is a small number and
\[\tau=(c_1,c_2, \cdots, c_U).\]
We define $\rho_{\epsilon}$ as the density operator of the reduced
state of the system of $N$ particles obtained from the maximally
mixed state in the restriction of the universe:
\begin{align} \rho_{\epsilon}=  \mbox{Tr}_E (\rho_{{\cal U},\epsilon, \max}) ,\label{canprob1}
\end{align}
where $\mbox{Tr}_E(\cdot)$ traces out the environment from the
universe and
\begin{align} \rho_{{\cal U},\epsilon, \max}=  \mbox{Tr}_E \frac{1} {|{\cal T}_{{\cal
U},\epsilon}|}\sum_{\tau \in {\cal T}_{{\cal U},\epsilon}}
\ket{\tau}\bra{\tau}\label{maximallymixed}
\end{align}
is the density operator of the maximally mixed state in the
restriction of the universe. For $U\rightarrow \infty$, by the Law
of Large Numbers, the restriction ${\cal H}_{{\cal U},\epsilon}$
contains almost all the eigenstates of the full Hilbert space of
the universe and all the eigenstates in the strongly typical set
share the same empirical one-particle distribution $\{P_c\}$ up to
a deviation whose probability is upper bounded  by Hoeffding's
inequality:
\[\mbox{Pr}\left(\left|\frac{n_c(\tau)}{U}-P_c\right| \geq \epsilon\right)
\leq 2e^{-2U\epsilon^2}, \ \forall \tau \in {\cal T}_{{\cal
U},\epsilon}.\] The above inequality shows that the density
operator $\rho_{{\cal U},\epsilon, \max}$ asymptotically tends to
become permutation invariant. By the quantum de Finetti theorem
\cite{caves,konig}, any permutation-invariant (exchangeable)
density operator on $U$ subsystems has the property that its
$N$-body reduced state can be approximated by a mixture of
independent and identically distributed (i.i.d.) product states as
$U \rightarrow \infty$:
\begin{align} \rho_{\epsilon} &\xrightarrow[]{\text{in probability}}
\left(\sum_cP_c\ket{c}\bra{c}\right)^{\otimes N} \label{definetti}
\\ &=\sum_sP_c^{n_c(s)}\ket{s}\bra{s}=\rho_{can},\label{canprob2}
\end{align}
where the last equality means that we take its left hand side as
the definition of system's canonical density operator
$\rho_{can}$.

When the system consists of indistinguishable bosons, the
symmetrization postulate requires that the physical Hilbert space
is the symmetric subspace of the full Hilbert space. System's
Hamiltonian, being invariant under particle permutations,
preserves this symmetric subspace. The symmetric Hilbert subspace
${\cal H}_{\cal N}$ of system's Hilbert space ${\cal H}_{\cal
S}={\cal H}_{\cal C}^{\otimes N}$, is spanned by the complete set
of ''bosonic'' eigenstates $\{\ket{{n}}, \ {n} \in {\cal N}\},$
with
\begin{align}\ket{{n}}= \sum_{{s} \in {\cal
S}_{{n}}}\frac{\ket{{s}}}{\sqrt{W_{{n}}}}, \nonumber 
\end{align}
see \cite{church} for a tutorial on the topic.

The following family of Kraus operators operates the mapping from
${\cal H}_{\cal S}$ to ${\cal H}_{\cal N}$:
\begin{align} K_{{n},{s}}
=\left\{
\begin{array}{cc} \ket{{n}}\bra{{s}} , &
 \mbox{if} \ {s} \in {\cal S}_{n},\\
0,  &  \mbox{elsewhere}.
\end{array} \right.
\nonumber 
\end{align}
The mapping is complete because
\begin{align} \sum_{{n}} \sum_{{s} }K_{{n},{s}}^{\dagger}K_{{n},{s}} &=
\sum_{{s}}\ket{s}\bra{s}, \nonumber
\end{align}
therefore the above family of Kraus operators defines a CPTP
linear map from ${\cal H}_{\cal S}$ to ${\cal H}_{\cal N}$.  We
now obtain the ''bosonic'' canonical density operator
${\nu}_{can}$ by mapping ${\rho}_{can}$ onto the symmetric
subspace:
\begin{align}{\nu}_{can}&=
\sum_{{n}} \sum_{{s}}K_{{n},{s}}
\rho_{can}K_{{n},{s}}^{\dagger}\nonumber
\\ &=
\sum_{{n} } \sum_{{s} \in {\cal S}_{{n}}}\ket{{n}}\braket{{s}|
\rho_{can}|{s}}\bra{{n}}\nonumber
\\ &=\sum_{{n} }\sum_{{s} \in {\cal S}_{{n}}}P_c^{n_c}\ket{{n}}\bra{{n}}
\nonumber
\\ &=\sum_{{n} }P_n \ket{{n}}\bra{{n}},
\nonumber 
\end{align}
with
\begin{align}P_n= W_{{n}}\prod_{c}
P_c^{n_c}.
 \label{multinomial}
\end{align}

The above equality  shows that the canonical probability
distribution of the occupancy numbers is the multinomial
distribution, see also \cite{pnas,zupa} for the multinomial
distribution in statistical mechanics. In the context of
probability theory, the one-particle probability distribution
$\{P_c\}$ that equips the multinomial distribution is called the
{\em categorical} distribution or {\em generalized Bernoulli}
distribution, the trials being generalized (non-binary) Bernoulli
(independent) trials and the $|{\cal C}|$ categories being the
$|{\cal C}|$ eigenstates. Here we don't need the i.i.d. assumption
as it emerges in (\ref{definetti}) from two ingredients:
\begin{itemize}
\item  The universe is in a permutation-invariant (fully
symmetric) maximally mixed state on a typical subspace.
 \item The
number of particles of the universe goes to infinity.
\end{itemize}

We observe that the product $\prod_{c}n_c!$ contained in $W_n$ is
a factor of the denominator of (\ref{multinomial}). Due to its
presence, the distribution (\ref{multinomial}) of the occupancy
numbers is not simply proportional to the product of factors,
hence, as desired, the occupancy numbers are not independent
random variables. The product $\prod_{c}n_c!$ in the denominator
of (\ref{multinomial}) is the answer given by the multinomial
distribution to the issue raised by Pathria and Beale in 1.6  of
their textbook \cite{pathria} about the role that it should have
in system's statistical properties. However, we cannot agree with
the successive analysis carried out in 5.4 of \cite{pathria},
whose result is that the sought distribution is the Bose-Einstein
distribution because, as already mentioned in the introduction,
the Bose-Einstein distribution applies to a system of the grand
canonical ensemble, not to a system of the canonical ensemble.

\section{Boltzmann's categorical distribution}

The Boltzmannian case is obtained by taking as categorical
probability distribution the Boltzmann one-particle probability
distribution:
\begin{align} P_c= \frac{e^{-\beta \epsilon_c}}{Z},
\ \  Z=\sum_{c} e^{-\beta \epsilon_c}, \label{boltzmann}
\end{align}
where $\epsilon_c$ is the energy eigenvalue of eigenstate $c$ and
$\beta>0$ is a real scalar. With (\ref{boltzmann}), for the
probability distribution of microstates and for the probability
distribution of the occupancy numbers we get
\begin{align} P_s&= \frac{\prod_{i=1}^Ne^{-\beta \epsilon_{c_i}}}
{Z^N}= \frac{\prod_{c} e^{-\beta \epsilon_c
n_c(s)}}{\sum_{n}W_{{{n}}}\prod_{c} e^{-\beta n_c \epsilon_c}},
\label{boltzmannmany}
\end{align}
\begin{align} P_n&=  \frac{W_n\prod_{c} e^{-\beta \epsilon_c
n_c}}{\sum_{{n}}W_{{{n}}}\prod_{c} e^{-\beta n_c \epsilon_c}},
\nonumber 
\end{align}
where we have substituted the multinomial expansion of $Z^N$,
\begin{align} Z^N=\left(\sum_{c } e^{-\beta
\epsilon_c}\right)^N=\sum_{{n} }W_{{{n}}}\prod_{c} e^{-\beta n_c
\epsilon_c}. \nonumber 
\end{align}

For completeness, hereafter we conclusively rule out the
probability distribution of occupancy numbers proposed in
\cite{reif} for a system at the thermal state, which is
\begin{align}P_{n,\small{\text{reif}}}&=\frac{\prod_{c}
e^{-\beta \epsilon_c n_c}}{Z_N(\beta)},\label{reif}
\end{align}
with
\begin{align}Z_N(\beta)&=\sum_{n }\prod_{c} e^{-\beta \epsilon_c n_c}
\label{gottlieb} \\&  =\sum_{n}W_{n}^{-1}\sum_{s \in {\cal
S}_n}\prod_{c} e^{-\beta \epsilon_c n_c(s)}\nonumber \\
&=\sum_{s}W_{n(s)}^{-1}\prod_{c} e^{-\beta \epsilon_c n_c(s)}.
\label{borrmann}
\end{align}
Just to link the above partition function to a couple of papers
drawn from the the vast literature that acknowledges (\ref{reif}),
we mention that (\ref{gottlieb}) is reported in eqn. (14)
of\cite{gottlieb}, while (\ref{borrmann}) is reported in eqn. (5)
of \cite{borrmann}.

We observe that, if (\ref{boltzmannmany}) holds, then the mutually
exclusive $W_n$ microstates belonging to ${\cal S}_n$ are equally
probable. Assuming equiprobability of the $W_n$ microstates
belonging to ${\cal S}_n$, the distribution of microstates
obtained from (\ref{reif}) is
\begin{equation}\frac{P_{n(s),\small{\text{reif}}}}{W_{n(s)}}=\frac{\prod_{c}
e^{-\beta \epsilon_c
n_c(s)}}{W_{n(s)}Z_N(\beta)}.\label{reif2}\end{equation} Since
$W_{n(s)}Z_N(\beta)\neq Z^N$, we see that (\ref{reif2}) is not
compatible with (\ref{boltzmannmany}), which not only descends
from (\ref{definetti}), but also is how is commonly understood a
canonical Boltzmannian system.

\section{Bosonic canonical typicality}

In \cite{spalvmacro} we obtained ${\nu}_{can}$ by tracing out the
environment from the universe in the case where, according to the
eigenstate thermalization hypothesis of \cite{eth}, the state of
the universe is the ''thermal'' bosonic eigenstate. In other
words, our previous result refers to the case in which the set
${\cal T}_{{\cal U},\epsilon}$ contains only and all the
eigenstates with the same occupation numbers, or, in the language
of typicality, only one type. We now substantially broaden the
scope of our previous result by tracing out the environment from
any pure state uniformly picked from ${\cal H}_{{\cal
U},\epsilon}$.

Let the universe be in any pure state $\ket{\phi}$ and consider
system's density operator
\[\rho_{\phi}=\mbox{Tr}_E(\ket{\phi}\bra{\phi}).\]
Using Levy's lemma,
 paper \cite{typpopescu} proves that, for $\ket{\phi}$ randomly
 picked with uniform probability inside the restricted Hilbert space
 ${\cal H}_{{\cal U},\epsilon}$ of the universe and for any $\eta>0$,
\begin{align}
\hspace{-0.1cm}\Pr(\,\| \rho_{\phi}-\rho_{\epsilon}\|_1 \ge \eta
\,) &\le 2\exp\!(-C\,|{\cal T}_{{\cal U},\epsilon}|([\eta -
\mu_{\Delta \rho}]_+)^2),\label{PSW2}
\end{align}
where
\begin{align}|| \rho||_1 = \mbox{Tr}(\sqrt{\rho
\rho^{\dagger}}) \nonumber
\end{align}
is the trace norm of the operator $\rho$, $C$ is a constant coming
from Levy's lemma,
\[[\eta - \mu_{\Delta \rho}]_+=\max\{\eta - \mu_{\Delta
\rho},0\},\] and
\[\mu_{\Delta \rho}=E\{\| \rho_{\phi}-\rho_{\epsilon}\|_1\},\]
where $E\{\cdot\}$ is the expectation over the random variable
inside the curly brackets, in this case, the random $\phi$. The
nontrivial exponential decay is obtained for
\[\eta \geq \mu_{\Delta \rho}.\]
A conservative estimate of the value of $\eta$ at which the decay
starts to be exponential is the left hand side of
\begin{equation}\sqrt{|{\cal S}| / d_E^{\rm eff}}\geq \mu_{\Delta
\rho},\nonumber \end{equation} where $d_E^{\rm
eff}=1/\mbox{Tr}((\mbox{Tr}_S(\rho_{{\cal U},\epsilon,\max}))^2)$
is the effective dimension of the environment and the inequality
is the main result of \cite{typpopescu}. For $U \gg N$, $d_E^{\rm
eff}$ is much bigger than $|{\cal S}|$ and values of $\eta$ much
greater than the very small number $\sqrt{|{\cal S}| / d_E^{\rm
eff}}$ are of interest. In this case, the bias term $\mu_{\Delta
\rho}$ inside the exponent of (\ref{PSW2}) can safely be
neglected.

Now let \(|{\cal N}_{{\cal U},\epsilon}|\) denote the number of
distinct occupation vectors (types) inside the restriction of the
universe.
 Since all microstates
belonging to the same occupation class are mapped onto the same
occupancy vector, the trace norm of the difference between the two
density operators in the system symmetric subspace depends only on
the $|{\cal N}_{{\cal U},\epsilon}|$ occupation vectors.
Therefore, applying Levy's lemma with dimension $|{\cal N}_{{\cal
U},\epsilon}|$ equal to the dimension of the restricted symmetric
subspace of the universe, the tail bound becomes
\begin{equation}\label{eq:refined-bound}
\hspace{-0.0cm}\Pr(\| \nu_{\phi}-\nu_{\epsilon}\|_1 \ge \eta) \le
2\exp(-C|{\cal N}_{{\cal U},\epsilon}|([\eta-\mu_{\Delta
\nu}]_+)^2).\hspace{-0.05cm}
\end{equation}
By the same arguments of \cite{typpopescu} it can be proved that
the upper bound above the bias becomes
\[\sqrt{|{\cal N}|/ d_{E,sym}^{\rm eff}}\geq \mu_{\Delta \nu},\]
where $d_{E,sym}^{\rm eff}=1/\mbox{Tr}((\mbox{Tr}_S(\nu_{{\cal
U},\epsilon,\max}))^2)$ is the effective dimension of the
environment in the symmetric subspace.

By typicality arguments it can be shown that
\begin{equation}|{\cal T}_{{\cal U},\epsilon}|\sim 2^{-U \sum_cP_c
\log_2(P_c)}.\label{typsize}
\end{equation}
At the same time,
\[|{\cal N}_{{\cal U},\epsilon}|\leq
\frac{(U+|{\cal C}|-1)!}{U!(|{\cal C}|-1)!}\]  scales only
polynomially with the number of particles of the universe, not
exponentially. Therefore, in the thermodynamic regime of the
universe, in which \(|{\cal N}_{{\cal U},\epsilon}|\ll |{\cal
T}_{{\cal U},\epsilon}|,\) and for values of $\eta$ of practical
interest, the bound \eqref{eq:refined-bound} can be much tighter
than the original bound (\ref{PSW2}), which, by contractivity of
the CPTP map, remains valid in the symmetric subspace as a
conservative estimate.

\section{Entropy of the multinomial distribution}

The Shannon entropy of the occupancy numbers, call it $H_n$, which
is equal to the von-Neumann entropy of $\nu_{can}$, is
\begin{align}
\hspace{-0.0cm} H_n &= - E\{\log(P_{{n}})\} \nonumber
\\ &= -NE\{\log(P_c)\}- E\{\log(W_{{n}})\}\label{multentr1}
\\ &=NH_c-\log(N!)+\sum_{c}E\{\log(n_c!)\}, \label{multentr}
\end{align}
where $H_c$ is the Shannon entropy of the categorical distribution
and $E\{\cdot\}$ denotes the expectation over the probability
distribution of the random variable appearing inside the curly
brackets. Specifically, since $n_c$ is a binomial random variable
$(N,P_c)$ we write
\begin{align}\hspace{-0.25cm}E\{\log(n_c!)\}&=\sum_{n_c=0}^NP_{n_c}\log(n_c!)
\nonumber \\ &= \sum_{n_c=0}^N\left(
\begin{array}{c}  N \\
 n_c
\end{array} \right)P_c^{n_c}(1-P_c)^{N-n_c}\log(n_c!)
\nonumber \\ &= \sum_{n_c=2}^N\left(
\begin{array}{c}  N \\
 n_c
\end{array} \right)P_c^{n_c}(1-P_c)^{N-n_c}\log(n_c!), \hspace{-0.15cm}
\label{difficult}\end{align} where
\[\left(
\begin{array}{c}  N \\
 n
\end{array} \right)=\frac{N!}{(N-n)!n!}\]
is the binomial coefficient.

Looking at (\ref{multentr})  we immediately recognize that $NH_c$
is the Shannon-Gibbs entropy of microstates and that $-\log(N!)$
is the Gibbsian correction term. The novelty introduced by
(\ref{multentr}) is the sum
\begin{align}&\sum_{c }E\{\log(n_c!)\},
\label{difficultsum}\end{align} that, added to the other terms,
makes $H_n$ a legal Shannon entropy and guarantees $H_n \geq 0$
also in the quantum regime.  The computational burden of the
numerical evaluation of this term can be mitigated as suggested in
\cite{me,mahdi}.

\section{Entropy of the thermal state}
In this section we study the entropy of a canonical system at the
thermal state and find the temperature of the system. The first
step is to work out the categorical distribution of the
multinomial distribution. Since the thermal state is characterized
by the constraint of expected energy, the categorical distribution
could be worked out by entropy maximization subject to the
constraint of expected energy. In general, due to the presence of
the term (\ref{difficultsum}), the one-particle Boltzmann
distribution, that comes out from maximization of the entropy of
microstates with expected energy constraint, will not be the
sought categorical maximizer. Constrained maximization is based on
the Lagrangian
\begin{align}
\hspace{-0.3cm} {\cal
L}=H_n+N\alpha\left(1-\sum_cP_c\right)+N\beta\left(\mu_{\cal
E}-\sum_cP_c \epsilon_c\right)\hspace{-0.1cm},\hspace{-0.2cm}
\label{lagrangian}
\end{align}
where the Lagrange multipliers $\alpha$ and $\beta$ impose the
constraints on the categorical distribution, specifically on
particle's expected energy
\begin{align}
\hspace{-0.0cm} \sum_{c }P_c\epsilon_c= \mu_{\cal E}
\label{energyconstraint}
\end{align}
and the obvious constraint
\begin{align}
\hspace{-0.0cm} \sum_{c }P_c=1, \label{probconstraint}
\end{align}
with the further constraint
\[P_c \geq 0, \ c=1,2 \cdots, |{\cal C}|.\]
 Unfortunately, due to the sum
(\ref{difficultsum}), excepting special cases (for instance the
case $N=1$ and the case where all the eigenstates have the same
energy), the entropy of the multinomial distribution is
intractable, making impossible to find the maximizer in closed
form.

To find a closed-form approximation to the actual maximizer, we
can plug into the Lagrangian a tractable approximation to the
actual intractable entropy. Let us consider the disjoint partition
of ${\cal C}$ into the subset ${\cal C}_w$ of well-populated
categories, say those categories that have non-negligible
probability of being occupied by two or more particles, and the
subset ${\cal C}_s$ of sparsely populated categories, and write
the entropy of the multinomial distribution as
\begin{align}
H_{n} = C_{s}+C_{w}+\log(N!), \label{sumcontr}
\end{align}
where $\log(N!)$ is independent of the categorical distribution
and therefore does not contribute to maximization and the
contribution $C_{s(w)}$ of the sparsely populated (well-populated)
categories belonging to ${\cal C}_{s(w)}$ is
\begin{align}
C_{s(w)} = \sum_{c \in {\cal C}_{s(w)}}E\{\log(n_c!)\}- NP_c
\log(P_c).\label{contributions}
\end{align}

The log-factorials of the sparsely populated categories are
negligible, therefore we  approximate
\begin{align}
C_{s} \approx - N\sum_{c \in {\cal C}_s}P_c \log(P_c). \label{cp}
\end{align}

The log-factorials of the well-populated categories are
approximated by truncating Stirling's asymptotic
expansion\footnote{Stirling's expansion is asymptotic, so we
cannot use it for the sparsely populated categories. At the same
time, we cannot neglect the log factorials of the well-populated
categories. To properly treat the two cases, we decided to split
the set of categories into well-populated and sparsely populated.}
to the first three terms of (\ref{stirling}):
\begin{align}
E\{\log(n_c!)\} \approx E\left\{n_c\log(n_c)-n_c+\frac{1}{2}\log(2
\pi n_c)\right\}.\label{threeterms}
\end{align}
For large $NP_c$ we further approximate
\begin{equation}\log(n_c)=\log(NP_c(1+v_c))
\approx \log(NP_c)+v_c-0.5v_c^2.\label{logapprox}
\end{equation}
where $v_c$ is a random variable with zero mean value and
\[E\{v^2\}=\frac{1-P_c}{NP_c}, \ \ E\{v^n\}={\cal O}(N^{1-n}) \ n>2.\]
 Neglecting the terms ${\cal O}(N^{-1})$, for the three
terms of (\ref{threeterms}) we get
\begin{align}
E\left\{n_c\log(n_c)\right\} \approx
NP_c\log(NP_c)+0.5(1-P_c),\label{first}
\end{align}\begin{align}
E\{n_c\}=NP_c,\label{second}
\end{align}\begin{align}
E\{\log(2 \pi n_c)\} \approx \log(2 \pi NP_c),\label{third}
\end{align}
which, inserted in (\ref{threeterms}) and then in
(\ref{contributions}), provide us with
\begin{align}
C_{w} = \sum_{c \in {\cal C}_{w}}NP_c(
\log(N)-1)+\frac{1-P_c+\log(2 \pi N P_c)}{2}.\label{cw}
\end{align}

 Plugging the sum of
(\ref{cp}) and (\ref{cw}) in place of $H_n$ in the Lagrangian and
putting to zero the partial derivatives w.r.t. $\{P_c\}$ we get
\begin{align} P_c=
\left\{
\begin{array}{cc}  \frac{1}{1+2N(\alpha+\beta \epsilon_c+1-\log(N))}, &
 \mbox{if} \ c \in {\cal C}_w,\\
e^{-1-\alpha-\beta \epsilon_c}, & \mbox{if} \ c \in {\cal C}_s,
\end{array} \right.
\label{spalvmax}\end{align} where $\alpha$ and $\beta$ can be
found numerically from (\ref{probconstraint}) and
(\ref{energyconstraint}). We see observe the probability
distribution of microstates with the above maximizer is not of
exponential type, or, in other words, it is not the many-particle
Boltzmann distribution. As in the standard analysis, it is always
equal to the product of factors but, in the general case, these
factors are not the Boltzmann factors. One unexpected consequence
of the non-exponential nature of (\ref{spalvmax}) is that the
probability of a microstate can be different from the probability
of another microstate even when the two have the same energy.

It is worth considering the two extreme cases, namely $C_w=0$ and
$C_s=0$. When $C_w=0$ all the categories are sparsely populated we
get rid of the entire sum (\ref{difficultsum}), leading to
\begin{equation}H_n \approx NH_c-\log(N!).
\label{classicalh}
\end{equation}
The categorical constrained maximizer of the right hand side is
the Boltzmann one-particle probability distribution, leading to
\begin{align}
\hspace{-0.25cm} H_{c}&= \beta \mu_{\cal E}+ \log(Z).
\label{classicalht}
\end{align}

When all the categories are well-populated, the result of the sum
(\ref{cw}) is
\begin{align}
C_{w} = N (\log(N)-1)+\frac{|{\cal C}|-1}{2}+\sum_{c}\frac{\log(2
\pi N P_c)}{2}.\label{cw2}
\end{align}
By substituting the first three terms of Stirling's expansion of
$\log(N!)$, for the entropy we find
\begin{align}
H_n&=C_{w}-\log(N!)\nonumber \\
&\approx \frac{1}{2}\left((|{\cal C}|-1)\log(2 \pi e
N)+\sum_{c}\log(P_c)\right),\label{ciapprw}
\end{align}
that is the leading term of the asymptotic expansion of
\cite{cichon}.


Note that the quality of the approximation will strongly depend on
how the set ${\cal C}$ of categories is partitioned. We approached
the issue by operating many different partitions. For each
partition we obtain a different categorical distribution, compute
the entropy of each one of the associated multinomial
distributions, and take as approximation to the actual maximizer
the categorical distribution leading to the highest entropy.

When an infinitesimally small reversible transformation at
constant density (number of particles over volume) is considered,
the temperature $T$, expressed in Kelvin degrees, of heat bath and
system at the equilibrium with it, is defined by plugging the
entropy of the occupancy numbers into Clausius' equation
\begin{align}\frac{d H_{n}}{Nd \mu_{\cal
E}}=\frac{d H_{c}}{d \mu_{\cal E}}+\frac{\sum_{c}d
E\{\log(n_c!)\}}{Nd \mu_{\cal E}}=\frac{1}{T},\label{clausiusmult}
\end{align}
where both entropy $H_{n}$ and expected system's internal energy
$N \mu_{\cal E}$, the quantity denoted $U$ in many textbooks, are
expressed here in $k$ units,
\[k=1.38 \cdot 10^{-23}\]
being the non-dimensional version of Boltzmann's constant.
 In practice, due to the difficulties
arising with the sum, the derivative of the entropy can be
approximated numerically as
\begin{align}\frac{d H_{n}}{d \mu_{\cal
E}} \approx \frac{[H_n]_{\mu_{\cal E}+\Delta}-[H_n]_{\mu_{\cal
E}-\Delta}} {2\Delta}.
\nonumber 
\end{align}
If the one-particle Boltzmann distribution is forced as
categorical distribution, then we have
\[\beta+\frac{\sum_{c}d
E\{\log(n_c!)\}}{N d \mu_{\cal E}}=\frac{1}{T}.\] Note that the
sum, which depends on $\mu_{\cal E}$ and on $\beta$, can make the
temperature substantially different from $1/\beta$ in certain
situations, for instance when a many-particle system in the
quantum regime is considered.

\section{Ideal quantum gas}
 In the case of a non-relativistic monoatomic
particle in a D-dimensional box with all the sides of length $L$,
the energy eigenvalues expressed in $k$-units with aperiodic
boundary conditions are
\begin{equation}\epsilon_{i_1,i_2, \cdots, i_D}= \sum_{d=1}^D\frac{h^2i_d^2}{8 k m
L^2},  \ \ i_d=0,1,\cdots, \hspace{-0.3cm}\label{boundaryd}
\end{equation}
where $m$ is the mass of the particle and
\[h=6.626 \cdot 10^{-34}\ \mbox{J}\cdot \mbox{s}\] is
Planck's constant.

In the classical regime, not only we can get rid of the sum
(\ref{difficultsum}), but also we can approximate the Boltzmann
probability distribution to a probability density function,
leading to
\begin{align}
\mu_{\cal E} &\approx \frac{DT}{2}, \label{classicalenergyd}
\end{align}
\begin{align}
\hspace{-0.4cm}H_n &\approx \frac{ND}{2}\left(\log \left(\frac{2
\pi k e m L^2}{\beta h^2}\right)\right) -\log(N!)
\label{twoterms} \\
& \approx N\log \left(\frac{eL^D}{Nh^D}\sqrt{\left(\frac{2 \pi k e
m}{\beta} \right)^D}\right), \label{st}
\end{align}
where the second approximation is obtained by substituting the
first two terms of the large-$N$ approximation (\ref{stirling}) of
to $\log(N!)$. As expected, plugging $\beta=1/T$ in (\ref{st}) we
get the semi-classical entropy formula originally derived by
Sackur and Tetrode by quantizing system's phase space into cells
of volume $h^{ND}$.

In the following, we will present numerical results obtained by
our approach for canonical systems and by the Bose-Einstein
approach for grand-canonical systems, therefore, for completeness,
we recall here also the formulas of Einstein's analysis of the
massive quantum gas that we use to derive the numerical results.
While in the semi-classical Einstein's analysis quantum states are
cells of volume $h^D$ in phase space, here we adopt the modern way
of expressing Einstein's results, which is based on the energy
eigenvalues (\ref{boundaryd}), see e.g.
\cite{kardar}.\footnote{Although the derivation of the
Bose-Einstein statistics from the energy eigenvalues obtained by
imposing the boundary conditions is a standard procedure adopted
in all the modern textbooks, we cannot refrain to observe that
imposing boundary conditions to a system of the grand-canonical
ensemble, that is a system whose particles, e.g. photons in the
case of the black-body, can join and leave it, sounds odd.
However, grand-canonical systems are out of the scope of this
research, which is limited to canonical systems. We leave this
observation as a basis for future investigations, adopting by now
the standard approach as it is.} With system's expected number of
particles $\mu_{N}$ and expected energy per particle $\mu_{\cal
E}$ as constraints, the entropy-maximizing distribution for
grand-canonical systems can be worked out exactly in closed form.
It is the product of geometric distributions, one for each quantum
eigenstate. The expected number of particles in a quantum
eigenstate with energy eigenvalue $\epsilon$ is the celebrated
Bose-Einstein statistics:
\[\mu_{\epsilon}=\frac{1}{e^{\beta \epsilon+\gamma}-1}.\]
The two Lagrange multipliers $\beta$ and $\gamma$ are found by
solving the system made by the two equations
\begin{equation}\mu_N=\sum_{\epsilon }\frac{m_{\epsilon}}{e^{\beta
\epsilon+\gamma}-1},\label{einsteinn}\end{equation}
\begin{align}\mu_N\mu_{\cal E}&  =\sum_{\epsilon }\frac{m_{\epsilon}\epsilon}{e^{\beta \epsilon+\gamma}-1}
,\label{einsteinenergy}\end{align}
 where the sum
is over the set of the energy eigenvalues (\ref{boundaryd}) and
$m_{\epsilon}$ is the multiplicity of the energy eigenvalue
$\epsilon$, i.e. the number of $D$-tuples $(i_1,i_2, \cdots, i_D)$
that, substituted in (\ref{boundaryd}), give $\epsilon$ as a
result. The entropy of the occupancy numbers is
\begin{align}H_{n}=\sum_{\epsilon}m_{\epsilon}\left(\frac{\beta \epsilon+\gamma}{e^{\beta
\epsilon+\gamma}-1}-\log(1-e^{-\beta \epsilon-\gamma})
\right).\label{einsteinentropy}\end{align} After the imposition of
Clausius' equation, the Lagrange multiplier $\beta$ turns out to
be equal to $T^{-1}$ (in this case, this happens also in the
quantum regime), and $-\gamma k T$ is what today we call the {\em
chemical potential}.

In the classical regime, it is standard to replace sums by
integrals and the multiplicity $m_{\epsilon}$ by the density of
states, leading again for expected energy and entropy to
(\ref{classicalenergyd}) and (\ref{st}), but with $\mu_N$ in place
of $N$. Note that, as the multinomial distribution, also the
Bose-Einstein approach brings by itself the Gibbsian correction
term in the entropy formula.

\section{Conclusive discussion}

In this section we present our conclusive comments. In doing this,
we help ourselves with numerical results obtained, for the
Bose-Einstein approach and for our approach, with particles of
mass $1.6735575 \cdot 10^{-27}$ kg in a 1D box of length $10^{-7}$
m. The expected number of particles is always equal to 100. In the
canonical case, the number of particles can formally be seen as a
random variable with mean value equal to 100 and with delta-type
distribution, so both the approaches meet the same constraint on
the expected number of particles and we expect that, with the same
expected energy, the entropy obtained from the Bose-Einstein
approach will always be not lower than the entropy of our
approach. We compare the Bose-Einstein entropy to ours not because
we want to compare the entropy of a grand-canonical system to the
entropy of a canonical system, rather because we want to compare
the entropy of the multinomial distribution with the
entropy-maximizing distribution with the same constraints.
\begin{figure}[!h]
\vspace*{-.3cm}
    \centering
    \hspace*{-.7cm}
    \includegraphics[width=.55\textwidth]{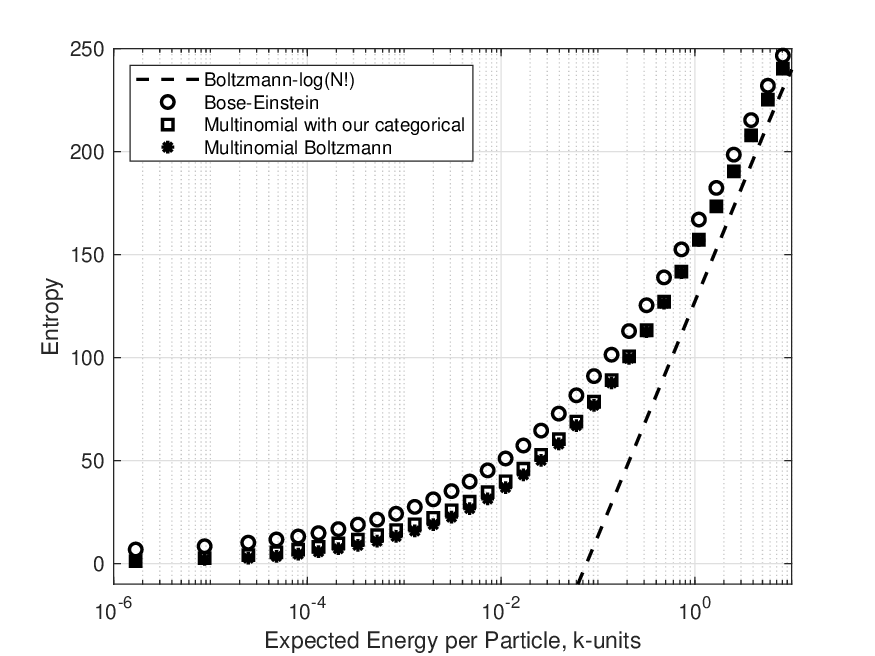}
    \caption{Various entropies in $k$ units versus expected energy per particle in $k$ units.
    The circles mark the entropy of the product of
    multinomial distributions obtained from Einstein's
    constrained entropy maximization.
    The squares mark the entropy of the multinomial distribution with
    optimized categorical distribution.
    The asterisks mark the entropy of the multinomial-Boltzmann
    distribution. The dashed line is the entropy
    of the Boltzmann distribution $-\log(N!)$.
}
    \label{fig:maxent}
    \vspace*{0.3cm}
\end{figure}

Fig. \ref{fig:maxent} reports the numerical results of entropy
versus expected energy per particle obtained by the product of
geometric distributions and by the multinomial distribution with
Boltzmann categorical distribution and with our proposed
categorical distribution. As expected, the entropy obtained by
Bose-Einstein approach is greater than the entropy of the
multinomial distribution. We observe that, for vanishingly small
expected energy, the Bose-Einstein entropy does not vanish.
Actually, all the particles tend to occupy the ground state, but
the randomness of their number is such that system's entropy tends
to be equal to the entropy of a geometrically distributed random
variable with mean value $\mu_N$.
Entropy becomes vanishingly small as the energy tends to zero if,
as many authors do, one passes from sums over energy
eigenstates
to integrals, see for instance chapter 7 of \cite{kardar}, hence
if one passes from discrete energy eigenvalues to continuous
energies. Actually, with dense energy, for $T>0$ the number of
distinct quantum states accessible to a particle is infinite, and,
as a consequence, the expected number of particles that occupy any
quantum state, including the ground state, is infinitesimally
small. This prevents the phenomenon described above that leads to
non-zero entropy also at arbitrarily small temperature.
However, it remains that, to our opinion, a fair comparison
between entropies versus energy must be made between systems that
have the same support set of energy eigenvalues, be the number of
particles random or fixed. Regarding the results obtained with the
multinomial distribution, we see that, as expected, the entropy
obtained with our proposed categorical distribution is not lower
than the entropy obtained with Boltzmann's categorical
distribution, the difference between the two being appreciable
only in the passage between the deep quantum regime and the
classical regime.

\begin{figure}[!h]
\vspace*{-.3cm}
    \centering
    \hspace*{-.7cm}
    \includegraphics[width=.55\textwidth]{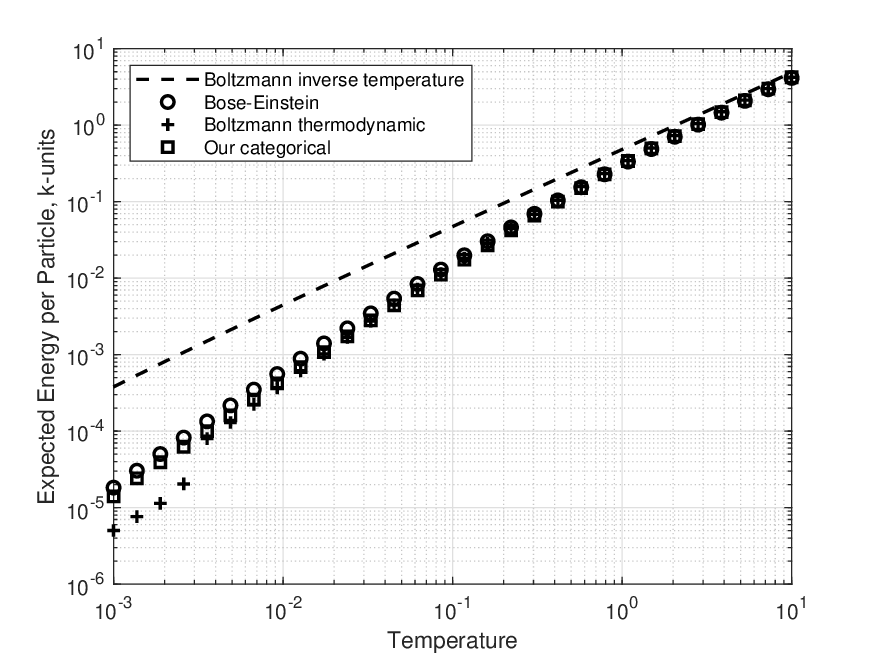}
    \caption{Expected energy per particle in $k$ units vs. temperature in various cases.
    The dashed line is the energy of the Boltzmann
    distribution with
    $\beta=1/T$. The circles mark the energy of the Bose-Einstein
    statistics.The squares mark the
    energy of our proposed categorical distribution and temperature obtained from (\ref{clausiusmult}). The symbols $+$ mark the
    energy of the Boltzmann
    distribution and temperature obtained from (\ref{clausiusmult}).}
    \label{fig:energyvstemp}
    \vspace*{0.0cm}
\end{figure}

\begin{figure}[!ht]
\vspace*{-.3cm}
    \centering
    \hspace*{-.7cm}
    \includegraphics[width=.55\textwidth]{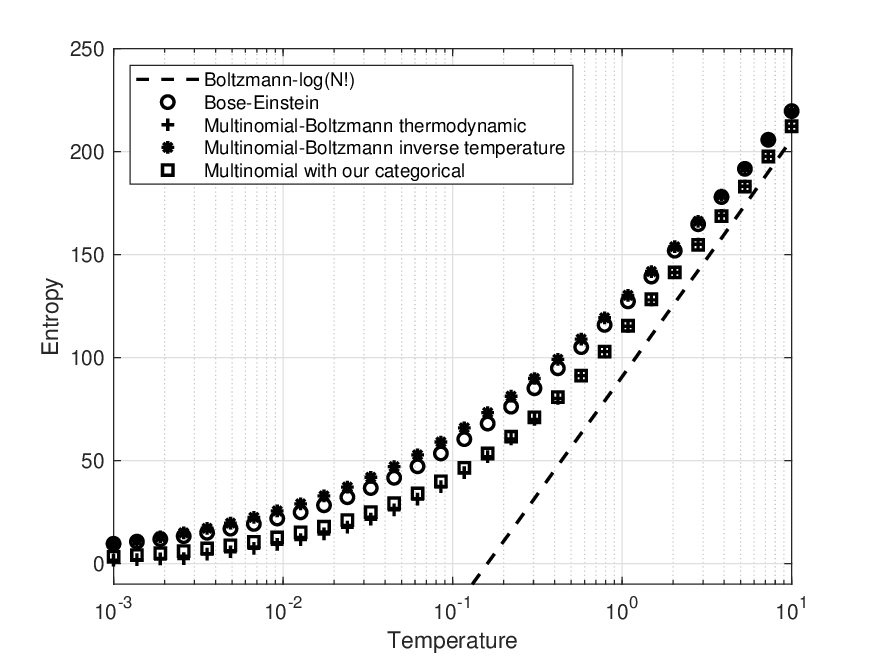}
    \caption{Various entropies in $k$ units versus temperature.
    The dashed line is the entropy of the Boltzmann distribution $-\log(N!)$,
    with $\beta=1/T$.
    The circles mark the Bose-Einstein entropy (\ref{einsteinentropy}).
    The asterisks mark the entropy of the Multinomial-Boltzmann
    distribution with $\beta=1/T$.
    The squares mark the entropy of the Multinomial distribution
    with our proposed categorical distribution and temperature obtained by imposing (\ref{clausiusmult}).
    The $+$ symbols mark the entropy of the Multinomial-Boltzmann distribution
    and temperature obtained by imposing (\ref{clausiusmult}). }
    \label{fig:thermoentropy}
    \vspace*{-.0cm}
\end{figure}

\begin{figure}[!h]
\vspace*{-.3cm}
    \centering
    \hspace*{-.7cm}
    \includegraphics[width=.55\textwidth]{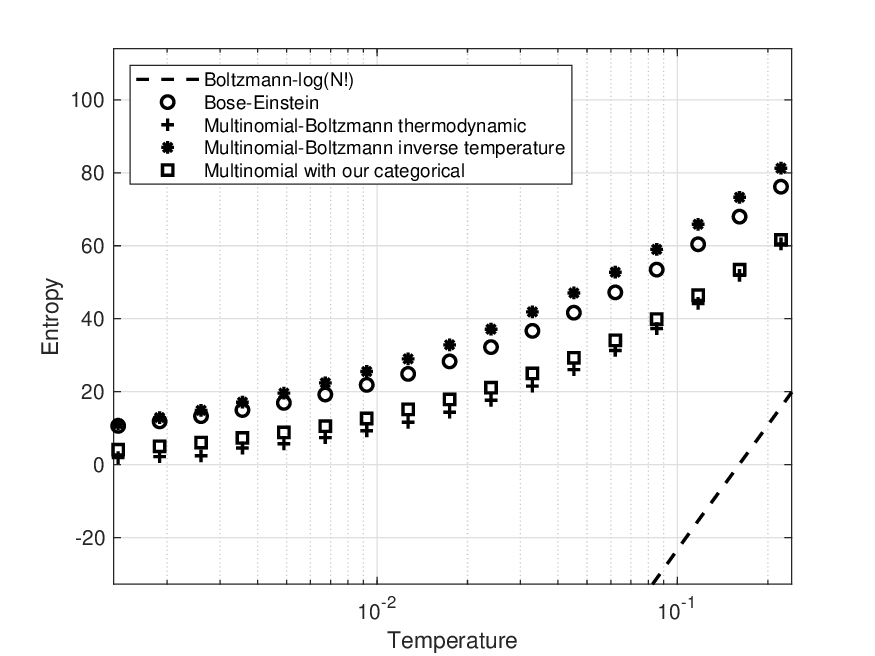}
    \caption{A detail of Fig. \ref{fig:thermoentropy},
    see the caption to
    Fig. \ref{fig:thermoentropy}.}
    \label{detail}
    \vspace*{-.3cm}
\end{figure}

When the temperature steps into the scene and we move our
attention to the entropy-temperature relation, the picture can
change because the energy-temperature relation depends on the
specific probability distribution at hand, so it cannot be
guaranteed that the probability distribution obtained from the
maximum entropy principle with energy constraint leads to maximum
entropy given the temperature. Fig. \ref{fig:energyvstemp} reports
the temperature-energy relation in various cases. This figure
shows that the energy of the Boltzmann distribution with $\beta$
equal to the inverse temperature is greater than the energy of the
Boltzmann distribution at the temperature obtained from eqn.
(\ref{clausiusmult}), and that it is also greater than the energy
obtained from the Bose-Einstein approach. To see the effect of the
change of abscissa from energy to temperature, we report various
entropies versus temperature in Fig. \ref{fig:thermoentropy}, a
detail of which is reproduced in Fig. \ref{detail}. A fact
strongly stands out from Fig. \ref{detail}: if we plug $\beta=1/T$
in the multinomial-Boltzmann distribution, then the resulting
entropy is greater than the Bose-Einstein entropy at the same
temperature, but the latter is obtained from the principle of
constrained entropy maximization. In other words, although, as it
must be and as it is clear from Fig. \ref{fig:maxent}, with the
same expected energy and expected number of particles the entropy
of the product of geometric distributions obtained from the
Bose-Einstein approach is greater than the entropy of the
multinomial-Boltzmann distribution, it happens that, plugging
$\beta=1/T$ in the multinomial-Boltzmann distribution leads to an
entropy that is not compatible with the principle of constrained
entropy maximization.
The reason is that, as it appears from Fig.
\ref{fig:energyvstemp}, the energy of the Boltzmann distribution
with $\beta=1/T$ is greater than the energy of the product of
geometric distributions at the same temperature and, as a
consequence, the entropy of the multinomial-Boltzmann distribution
can potentially be greater than the entropy of the product of
geometric distributions at the same temperature. However, the
results reported in Fig. \ref{fig:energyvstemp} obtained with
temperature derived by imposing Clausius' equation save, at least
in this example, the coexistence between the maximum entropy
principle, and show a little advantage of our proposed categorical
distribution over Boltzmann's categorical distribution.

Before the conclusion of this paper, we comment one more time a
result that, for sure, is unexpected by most readers. As it is
clear from the Gibbs paradox, from the universally recognized
success of the Bose-Einstein approach based on the occupancy
numbers of grand-canonical systems, and, in the end, from the fact
that thermodynamics deals with macroscopic properties of the
system, microstates are not a thermodynamic entity. The entire
thermodynamics of systems at the equilibrium descends from the
statistical properties of the occupancy numbers and, for this
reason, there is no doubt that thermodynamics can survive without
the concept itself of microstates and without the equality
$\beta=1/T$ deriving from the imposition of Clausius' equation on
their entropy. If, due to affection for the equation $\beta=1/T$,
we want to keep microstates and their entropy among the
fundamental entities of thermodynamics, then Fig. \ref{detail}
forces us to renounce to the coexistence between the maximum
entropy principle and thermodynamics of canonical systems. But, to
our opinion, this renounce would be much more disruptive than the
renounce to the vision of microstates as a fundamental entity of
thermodynamics.

\end{document}